\documentclass[journal]{IEEEtran}

\usepackage{cite}      
\usepackage{graphicx}  
\usepackage{psfrag}    
\usepackage{subfigure}               

\usepackage{amsmath}   
\usepackage{latexsym}
\usepackage{amssymb}
\usepackage{bm}

\newcommand{\pdag}{{\phantom{\dagger}}}
\newcommand{\bq}{\begin{equation}}
\newcommand{\eq}{\end{equation}}
\newcommand{\bn}{\begin{eqnarray}}
\newcommand{\en}{\end{eqnarray}}

\begin{document}
\title{Robust negative differential conductance and enhanced shot noise in transport through a 
molecular transistor with vibration assistance}

\author{Bing Dong,
        X. L. Lei,
        and N. J. M. Horing,
\thanks{Bing Dong and X.L. Lei are with the Department of Physics, Shanghai Jiaotong University,
1954 Huashan Road, Shanghai 200030, China.}
\thanks{N.J.M. Horing is with the Department of Physics and Engineering Physics, Stevens 
Institute of Technology, Hoboken, New Jersey 07030, USA.}
}

\markboth{ID number 92 }%
{Shell \MakeLowercase{\textit{et al.}}: Bare Demo of IEEEtran.cls for Journals}

%



\maketitle

\begin{abstract}
In this paper, we analyze vibration-assisted sequential tunneling (including current-voltage 
characteristics and zero-frequency shot noise)
through a molecular quantum dot with two electronic orbitals asymmetrically coupled to the 
internal vibration. We employ rate equations for the case of equilibrated phonons, and strong 
Coulomb blockade. 
We find that a system with a strongly phonon-coupled ground state orbital and weakly 
phonon-coupled excited state orbital exhibits strong negative differential conductance; and it 
also shows super-Poissonian current noise. We discuss in detail the reasons and conditions for 
the appearance of negative differential conductance.
\end{abstract}


%
\IEEEpeerreviewmaketitle

\section{Introduction}

\PARstart{I}{n} a nanoelectromechanical system, mechanical motion can affect electrical transport 
and vice versa.\cite{Roukes} This has stimulated a large body of 
experimental\cite{jPark,hPark,Zhitenev,Yu,Pasupathy,LeRoy,Sapmaz} and 
theoretical\cite{Bose,Alexandrov,McCarthy,Mitra,Koch,Koch1,Koch2,Zazunov,Nowa
ck,Wegewijs,Haupt} work concerning phonon-assisted resonant tunneling through a quantum dot (QD) 
with strong coupling to an internal vibrational (phonon) mode (IVM), since tunneling of a single 
electron may induce a displacement of the mobile mechanical structure (which can be used as a 
prototype of a nanoscale displacement sensor). On the other hand, the vibration of an adsorbed 
molecule can induce a change of tunneling current, providing a mechanism for sensing its 
adsorption and its identification. Current progress in nanotechnology has facilitated the 
fabrication of single-electron tunneling devices using molecules, for instance, a single 
C$_{60(140)}$ molecule and carbon nanotubes (CNTs). In particular, CNTs are ideal systems for 
designing electromechanical devices since they have a small diameter, a low mass, hence a strong 
electron-phonon coupling (EPC) constant, and more importantly a relatively high quality factor, 
up to $10^4$.   

Recent experimental studies of a long suspended CNT sample connected to two electrodes has 
revealed the appearance of negative differential conductance (NDC) at the onset of each 
phonon-assisted current step.\cite{Sapmaz}
Theoretically, several previous calculations have been carried out for a system of a single level 
coupled to the phonon mode, and have revealed that the combined effect of low vibrational 
dissipation, i.e. an unequilibrated phonon (hot phonon), and strongly asymmetric tunnel-couplings 
to the left and right electrodes is responsible for the appearance of NDC at an appropriate EPC 
strength.\cite{Bose,Koch2,Zazunov,Shen} However, this NDC in the single-orbital system is quite 
weak and can be destroyed easily by a dissipative environment\cite{Shen}, and even by applying an 
ac-bias voltage.\cite{Dong} Nonetheless, it is desirable to search for a strong and robust NDC 
property in such transistors by tailoring material parameters for possible application in 
functional devices.   

Our recent work has examined unequilibrated vibration-assisted tunneling through a molecular QD 
with {\em two} electronic orbitals having asymmetric couplings to the electrodes, with both 
strongly interacting with an IVM.\cite{Shen} We found a strong NDC and correspondingly enhanced 
current noise due to a trapping effect, and we analyzed in detail the influence of the 
uneqilibrated vibration on its appearance. 
On the other hand, Nowack and Wegewijs have considered vibration-assisted tunneling through a 
two-level QD with asymmetric couplings to an IVM.\cite{Nowack} Their results display strong NDC 
through a competition between different Franck-Condon (FC) tailored tunneling processes of the 
two levels, which is completely robust even against strong relaxation of the vibration. In this 
paper, we further investigate tunneling in this two-orbital system with asymmetric couplings to 
the vibration in the equilibrated vibration condition, focusing on the reasons and conditions for 
the appearance of strong NDC, as well as its corresponding zero-frequency shot noise. We find 
that the current noise is always greatly enhanced (having a giant Fano factor) for this kind of 
system.

The outline of this paper is as follows. In Sec.~II, we describe the model system that we study. 
We also exhibit the rate equations in a number-resolved form at high temperature to describe 
phonon-assisted resonant tunneling in the limit of strong relaxation (equilibrated phonon).
In Sec.~III, we then investigate in detail the vibration-assisted transport and shot noise 
properties of a molecular two-orbital QD. Finally, a brief summary is given in Sec.~IV.

\section{Model and rate equations}


Here we consider a generic model, as shown in Fig.~1, for a molecular QD with two spinless 
levels, one as the highest-occupied molecular orbital (HOMO), $\varepsilon_1$, and the other as 
the lowest-unoccupied molecular orbital (LUMO), $\varepsilon_2$, coupled to two electrodes, left 
(L) and right (R), and also linearly coupled to an IVM of the molecule having frequency 
$\omega_0$ with respective coupling strengths $\lambda_1$ and $\lambda_2$. The model Hamiltonian 
is
\begin{subequations}
\bq
H = H_{leads}+H_{mol}+H_{T}, \label{hamiltonian}
\eq
with
\bn
H_{leads} &=& \sum_{\eta, {\bf k}} \varepsilon_{\eta {\bf k}} c_{\eta {\bf k}}^\dagger c_{\eta 
{\bf k}}^\pdag, \\
H_{mol} &=& \sum_{j=1,2} \varepsilon_j c_j^{\dagger} c_j^\pdag +U n_1 n_2 \cr
&& + \omega_0 a^\dagger a + \sum_{j=1,2} \lambda_j c_j^\dagger c_j^\pdag (a^\dagger + a), 
\label{Hmol} \\
H_T &=& \sum_{\eta,{\bf k},j} (V_{\eta j} c_{\eta {\bf k}}^\dagger c_j + {\rm H.c.}),
\en
\end{subequations}
where $c_{\eta{\bf k}}^\dagger$ ($c_{\eta{\bf k}}$) is the creation (annihilation) operator of an 
electron with momentum ${\bf k}$, and energy $\varepsilon_{\eta {\bf k}}$ in lead $\eta$ 
($\eta=L,R$), and $c_{j}^\dagger$ ($c_j$) is the corresponding operator for a spinless electron 
in the $j$th level of the QD ($j=1,2$). $U$ denotes interdot Coulomb interaction and 
$n_j=c_j^\dagger c_j^\pdag$ is the electron number operator in level $j$. In this paper, we 
consider strong Coulomb repulsion, $U\rightarrow \infty$, which suppresses simultaneous 
occupation of the two orbitals. $a^\dagger$ ($a$) is phonon creation (annihilation) operator for 
the IVM. $\lambda_j$ represents the EPC constant of an electron in level $j$; $V_{\eta j}$ 
describes the tunnel-coupling between electron level $j$ and lead $\eta$.
We use units with $\hbar=k_B=e=1$ throughout the paper. In addition, it is worth noting that we 
do not consider the EPC-induced coupling between the two MO's in the Hamiltonian, 
Eq.~(\ref{Hmol}), because it is much weaker than the coupling of a single MO. 

\begin{figure}[htb]
\centering
\includegraphics[width=6.5cm]{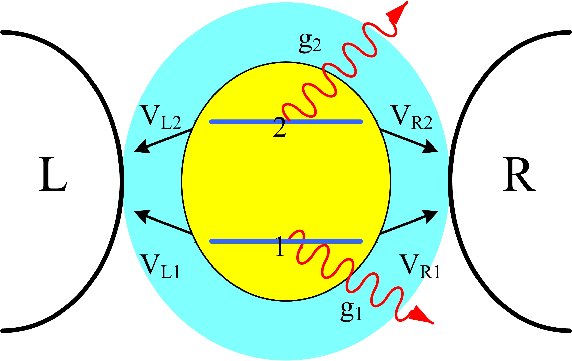}
\caption{Schematic diagram of the model system.}\label{FIG.1}
\end{figure}

It is well-known that the electron-phonon interaction term in Eq.~(\ref{Hmol}) can be eliminated 
by a canonical transformation,\cite{Mahan} $\widetilde{H}=e^{S} H e^{-S}$, with $S=(g_1 n_1 + g_2 
n_2) (a^\dagger - a)$ ($g_j=\lambda_j/\omega_0$), leading to a renormalization of the energy 
parameters, $\widetilde{\varepsilon}_j=\varepsilon_j-g_j\lambda_j$, and of the tunnel-coupling, 
$\widetilde{V}_{\eta j}=V_{\eta j} \exp [g_j(a^\dagger +a)]$. The transformed Hamiltonian is
\begin{subequations}
\bn
\widetilde{H}_{mol} &=& \sum_{j} \widetilde{\varepsilon}_j c_j^\dagger c_j + \widetilde{U} n_1 
n_2 + \omega_0 a^\dagger a, \\
\widetilde{H}_{T} &=& \sum_{\eta,{\bf k},j} (\widetilde{V}_{\eta j} c_{\eta {\bf k}}^\dagger c_j 
+ {\rm H.c.}),
\en
\end{subequations}
with $\widetilde{U} = U- 2\frac{\lambda_1 \lambda_2}{\omega_0}$.

In this paper, we consider the case of strong relaxation of the IVM. In this situation, the 
vibration is always in an equilibrated state during tunneling processes, i.e., the excited phonon 
relaxes very quickly due to strong dissipation, before the next electronic tunneling event takes 
place. As a result, in this strongly Coulomb-interacting two-level QD system, we have 
$\rho_{00}$, $\rho_{11}$, and $\rho_{22}$ denoting probabilities for the electronic states: 
zero-electrons in the dot ($0$), single-electron occupation in the HOMO ($1$) or the LUMO ($2$), 
respectively.

In the weak tunneling regime and high temperature approximation, $\Gamma_{\eta j} \ll T, 
\omega_0$ ($\Gamma_{\eta j}$ is the tunneling rate between lead $\eta$ and level $j$, and $T$ is 
the temperature), we have established generic rate equations for these electron occupation 
probabilities including the role of phonon-assisted resonant tunneling in the eqilibrated phonon 
condition employing a quantum Langevin equation approach.\cite{Shen} 
In order to employ MacDonald's formula for calculating shot noise\cite{MacDonald,Chen}, we write 
these rate equations in a number-resolved form: 
\begin{subequations}
\bn
\dot \rho_{00}^{(l)} &=& \Gamma_{L1}^- \rho_{11}^{(l)} + \Gamma_{L2}^- \rho_{22}^{(l)} + 
\Gamma_{R1}^- \rho_{11}^{(l-1)} + \Gamma_{R2}^- \rho_{22}^{(l-1)} \cr
&& - (\Gamma_{1}^+ + \Gamma_{2}^+) \rho_{00}^{(l)} , \label{qre:r00} \\
\dot \rho_{11}^{(l)} &=& \Gamma_{L1}^+ \rho_{00}^{(l)} + \Gamma_{R1}^+ \rho_{00}^{(l+1)}  - 
\Gamma_{1}^- \rho_{11}^{(l)}, \\
\dot \rho_{22}^{(l)} &=& \Gamma_{L2}^+ \rho_{00}^{(l)} + \Gamma_{R2}^+ \rho_{00}^{(l+1)} - 
\Gamma_{2}^- \rho_{22}^{(l)}, \label{qre:r22}
\en
\end{subequations}
where $\rho_{jj}^{(l)}(t)$ represents the electron occupation probability at time $t$ together 
with $l$ electrons arriving at the right lead due to tunneling events. Obviously, 
$\rho_{jj}(t)=\sum_{l} \rho_{jj}^{(l)}(t)$. 
The electronic tunneling rates are defined as
\begin{subequations}
\bn
\Gamma_{j}^+ &=& \sum_{\eta} \Gamma_{\eta j}^+= \sum_{\eta} \Gamma_{\eta j} \sum_{n} {\cal 
L}_{n}^j f_{\eta}(\widetilde {\epsilon}_j + n\omega_0), \cr
&& \\
\Gamma_{j}^- &=& \sum_{\eta} \Gamma_{\eta j}^- \cr
&=& \sum_{\eta} \Gamma_{\eta j} \sum_{n} {\cal L}_{n}^j [1-f_{\eta}(\widetilde {\epsilon}_j 
-n\omega_0)], \label{Lnj1} \\
{\cal L}_n^j &=& e^{-g_j^2(2n_B+1)} e^{n\omega_0/2T} {\cal I}_n(2g_j^2 \sqrt{n_B(n_B+1)}), 
\label{Lnj2}
\en
\end{subequations}
where $f_{\eta}(\epsilon)=[1+e^{(\epsilon-\mu_{\eta})/T}]^{-1}$ is the Fermi-distribution 
function of lead $\eta$ ($\mu_{\eta}$ is the chemical potential of lead $\eta$), 
$n_B=(e^{\omega_0/T}-1)^{-1}$ is the Bose distribution function and ${\cal I}_n(x)$ is the $n$th 
Bessel function of complex argument.

The current flowing through the system can be evaluated by the time rate of change of electron 
number in the right lead as
\bq
I = \dot N_{R}(t) =\frac{d}{dt} \sum_{l} l P^{(l)}(t) {\Big |}_{t\rightarrow\infty}, \label{Inr}
\eq
where
\bq
P^{(l)}(t) = \sum_{n} [\rho_{00}^{(l)}(t)+ \rho_{11}^{(l)}(t)+ \rho_{22}^{(l)}(t) ]
\eq
is the total probability of transferring $l$ electrons into the right lead by time $t$. The 
zero-frequency shot noise with respect to the right lead is similarly defined in terms of 
$P^{(l)}(t)$ as well:\cite{Shen,MacDonald,Chen}
\bq
S(0)=2\frac{d}{dt} \left [ \sum_{l} l^2 P^{(l)}(t) - (t I)^2 \right ] {\Big 
|}_{t\rightarrow\infty}. \label{snnr}
\eq

To evaluate $S(0)$, we define an auxiliary function $G_{jj}^{\eta }(t)$ as
\bq
G_{jj}(t) = \sum_{l} l \rho_{jj}^{(l)}(t),
\eq
whose equations of motion can be readily deduced employing the number-resolved QREs, 
Eqs~(\ref{qre:r00})--(\ref{qre:r22}), in matrix form: $\dot{\bm G}(t)={\cal M} {\bm G}(t) + {\cal 
G} {\bm \rho}(t)$ with ${\bm G}(t)=(G_{00}, G_{11}, G_{22})^{T}$ and ${\bm \rho}(t)=(\rho_{00},  
\rho_{11}, \rho_{22})^{T}$.  
Applying the Laplace transform to these equations yields
\bq
{\bm G}(s) = (s {\bm I}-{\cal M})^{-1} {\cal G} {\bm \rho}(s),
\eq
where ${\bm \rho}(s)$ is readily obtained by applying the Laplace transform to its equations of 
motion with the initial condition ${\bm \rho}(0)={\bm \rho}_{st}$ (${\bm \rho}_{st}$ denotes the 
stationary solution of the rate equations). Due to the inherent long-time stability of the 
physical system under consideration, all real parts of nonzero poles of ${\bm \rho}(s)$ and ${\bm 
G}^{\eta}(s)$ are negative definite. Consequently,  the divergent terms arising in the partial 
fraction expansions of ${\bm \rho}(s)$ and ${\bm G}^{\eta}(s)$ as $s\rightarrow 0$ entirely 
determine the large-$t$ behavior of the auxiliary functions, i.e. the zero-frequency shot noise, 
Eq.~(\ref{snnr}). 
This scheme facilitates evaluation of the zero-frequency shot noise as a function of bias-voltage 
for equilibrated-vibration-assisted tunneling through a molecular QD.

\section{Results and discussions}

In this section, we proceed with numerical calculations of the current $I$, the zero-frequency 
Fano factor $F=S(0)/2I$ for the two-orbital model. We set the symmetric tunnel-couplings as 
$\Gamma_{L1}=\Gamma_{R1}=\Gamma_{L2}=\Gamma_{R2}=0.01\omega_0$, take a weak EPC constant of the 
excited level, $g_2=0.1$, with the energy gap between the LUMO and HOMO as 
$\Delta=\widetilde{\varepsilon}_2-\widetilde{\varepsilon}_1= \omega_0/2$, in our calculations 
(Here, several values of $g_1$ are employed). 
Throughout the paper, the energy of IVM, $\omega_0$, is chosen as the energy unit, and we set the 
temperature as $T=0.05\omega_0$ unless otherwise indicated and assume that the bias voltage $V$ 
is applied symmetrically, $\mu_L=-\mu_R=V/2$. We have checked our parameter choices by comparing 
with the experimental data.\cite{hPark,Sapmaz} Our numerical calculations yield the maximum 
current through the molecule as $I_{\rm max} \sim 0.5 \Gamma$ (see below), and the experimental 
result is about $I_{\rm max} \sim 100$ pA, therefore we obtain $\Gamma \sim 1 \,\mu$eV, which is 
nearly $\sim 0.01\omega_0$ since the characteristic energy $\omega_0$ of the oscillation is found 
to be $\sim 0.5\,$meV for the experimental setup of Sapmaz, {\it et al.}\cite{Sapmaz}   

Figure 2 exhibits the calculated results. For comparison, we also plot the corresponding results 
of the single-level QD (which has only the ground orbital involved in transport) in Fig.~2(a,b). 
It is clearly evident that for the single-level system, the strong coupling between electronic 
and vibrational degrees of freedom induces a significant current suppression at low bias voltage 
due to a {\em phonon-modified tunneling rate}, and the noise is always sub-Poissonian, albeit the 
current is also largely suppressed by the strong EPC. Interestingly, it is seen that the current 
noise has an oscillatory feature with an approximate period of $V=2.0\omega_0$, if 
$\widetilde{\varepsilon}_1\neq \pm n\omega_0/2$ ($n$ is an integer), which can be explained based 
on Eqs.~(\ref{qre:r00})-(\ref{Lnj2}). For the sake of simplicity, we assume low temperature, 
$T=0$, in the following analysis. 

\begin{figure}[htb]
\centering
\includegraphics[height=8cm,width=8.5cm]{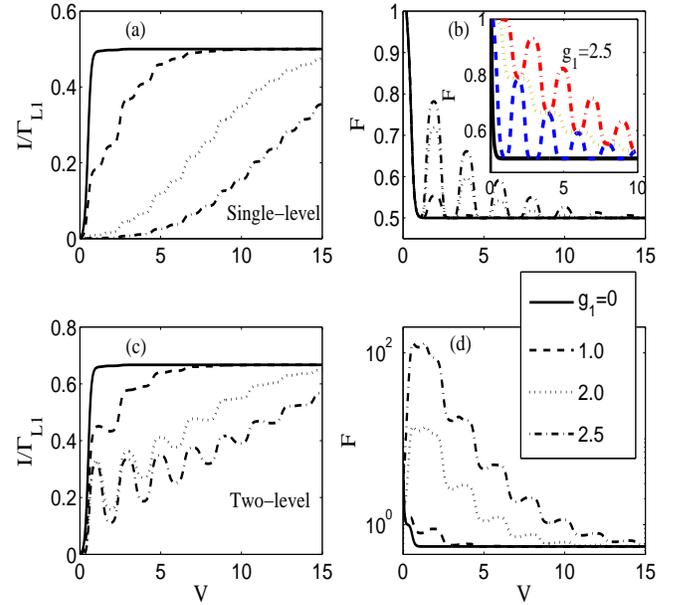}
\caption{Current $I$ (a,c) and Fano factor $F=S(0)/2I$ (b,d) vs. the bias voltage for a 
single-level system, $\widetilde\varepsilon_{1}/\omega_0=-0.25$ (a,b), and for a two-level 
system, $\widetilde\varepsilon_{1}/\omega_0=-0.25$, $\Delta/\omega_0=1/2$, $g_2=0.1$ (c,d), with 
various values of $g_1$. Inset: Fano factor vs. bias voltage for the single-level system with 
$g_1=2.5$ and various energies of the QD level, $\widetilde{\varepsilon}_1=0$ (solid line), 
$-0.25$ (dashed line), $-0.5$ (dotted line), and $-0.75$ (dotted-dashed line).}\label{FIG.2}
\end{figure}

To start, we consider a system with $|\widetilde{\varepsilon}_1|<\omega_0/2$, for example, 
$\widetilde{\varepsilon}_1=-0.25\omega_0$. Within the bias voltage window 
$0.5\omega_0<V<1.5\omega_0$, we have the tunneling-in rate, $x'=\Gamma_L^+=\Gamma_{L1} {\cal 
L}_{0}^1$, and the tunneling-out rate, $x=\Gamma_{R}^-=\Gamma_{R1} {\cal L}_{0}^1$ 
($\Gamma_{R}^+=\Gamma_{L}^-=0$). This is equivalent to a single-level QD with symmetric 
tunnel-couplings under large bias voltage. Its typical Fano factor is $F_1=1/2$. However, when 
the bias voltage becomes $1.5\omega_0<V<2.5\omega_0$, the $n=1$ channel is open, which leads to 
$x'=\Gamma_L^+=\Gamma_{L1} ({\cal L}_{0}^1 + {\cal L}_{1}^1)$ but still has 
$x=\Gamma_{R}^-=\Gamma_{R1} {\cal L}_{0}^1$. The equivalent model changes to an asymmetric 
single-level system with a Fano factor given by 
\bq
F_2=\frac{x'^2 + x^2}{(x' + x)^2}<1. \label{F2}  
\eq
If the bias voltage increases further to $2.5\omega_0<V<3.5\omega_0$, $\Gamma_R^-$ becomes equal 
to $\Gamma_{L}^+$ again and the Fano factor changes back to $1/2$. Therefore, because of the 
specific forms of the phonon-assisted tunneling rates, Eqs.~(\ref{Lnj1}) and (\ref{Lnj2}), the 
single-level QD varies between two equivalent models, the symmetric model and asymmetric one, 
with increasing bias voltage, which generates the oscillatory feature of the Fano factor. 
Similarly, it is easy to verify that a system with $\widetilde{\varepsilon}_1=0$ always 
corresponds to the symmetric model with a constant Fano factor, $F_1=1/2$.

A system with $|\widetilde{\varepsilon}_1|\geq \omega_0/2$ always corresponds to the asymmetric 
model, whose noise exhibits step structure only if $|\widetilde{\varepsilon}_1|=n\omega_0/2$ ($n$ 
is a positive integer), but it is oscillatory otherwise, as shown in the inset of Fig.~2(b). 
Here, for illustration, we take $\widetilde{\varepsilon}_1=-0.5\omega_0$ and $-0.75\omega_0$ as 
examples. Their zero-temperature tunneling rates are listed in Table~I for three different bias 
voltages. It is obvious that the rates of tunneling-in and -out, $x'$ and $x$, of the system with 
$\widetilde{\varepsilon}_1=-0.5\omega_0$ experience a simultaneous increase with increasing bias 
voltage, leading to a step-down of the current noise; but their increases become out of phase 
with respect to bias voltage for a system with $\widetilde{\varepsilon}_1=-0.75\omega_0$: the 
tunneling-in rate ($x'$) suffers an increase first and the tunneling-out rate ($x$) remains 
unchanged from $V=2.0\omega_0$ to $3.0\omega_0$, then $x$ increases at $V=4.0\omega_0$, which 
generates an oscillation of the noise [Eq.~(\ref{F2})].      

\begin{table*}[!t]
\renewcommand{\arraystretch}{1.3}
\caption{Tunneling rates, $x'=\Gamma_{L1}^+$ and $x=\Gamma_{R1}^-$, of the ground orbital with 
$\widetilde{\varepsilon}_1=-0.5\omega_0$ and $-0.75\omega_0$ at $V/\omega_0=2.0$, $3.0$, and 
$4.0$, respectively, at zero-temperature.}
\centering
\begin{tabular}{|c|c|c|c|c|}
\hline
& & $V/\omega_0=2.0$ & $3.0$ & $4.0$\\
\hline
$\widetilde{\varepsilon}_1=-0.5\omega_0$ & $x'$ & $\Gamma_{L1} ({\cal L}_{0}^1 + {\cal L}_{1}^1)$ 
& $\Gamma_{L1} ({\cal L}_{0}^1 + {\cal L}_{1}^1+ {\cal L}_{2}^1)$ & $\Gamma_{L1} ({\cal L}_{0}^1 
+ {\cal L}_{1}^1+ {\cal L}_{2}^1)$ \\
 & $x$ & $\Gamma_{L1} {\cal L}_{0}^1$ & $\Gamma_{L1} ({\cal L}_{0}^1 + {\cal L}_{1}^1)$ & 
$\Gamma_{L1} ({\cal L}_{0}^1 + {\cal L}_{1}^1)$ \\
\hline
$\widetilde{\varepsilon}_1=-0.75\omega_0$ & $x'$ & $\Gamma_{L1} ({\cal L}_{0}^1 + {\cal 
L}_{1}^1)$ & $\Gamma_{L1} ({\cal L}_{0}^1 + {\cal L}_{1}^1+ {\cal L}_{2}^1)$ & $\Gamma_{L1} 
({\cal L}_{0}^1 + {\cal L}_{1}^1+ {\cal L}_{2}^1)$ \\
 & $x$ & $\Gamma_{L1} {\cal L}_{0}^1$ & $\Gamma_{L1} {\cal L}_{0}^1$ & $\Gamma_{L1} ({\cal 
L}_{0}^1 + {\cal L}_{1}^1)$ \\
\hline
\end{tabular}
\end{table*}
 
In the case of a single-level QD, there is no NDC and no super-Poissonian noise. However, the 
situation is different for two-level systems. We find a strong NDC and a huge Fano factor ($F\sim 
100$ for $g_1=2.5$) with $\widetilde{\varepsilon}_1=-0.25\omega_0$, $\Delta=\omega_0/2$, and 
$g_1>1.0$, as shown in Figs.~2(c,d). Actually, we find from Fig.~3 that no NDC occurs only when 
$\widetilde{\varepsilon}_1= \pm n\omega_0/2$ between 
$-1.5\omega_0<\widetilde{\varepsilon}_1<1.0\omega_0$.   

\begin{figure}[htb]
\centering
\includegraphics[width=7.5cm]{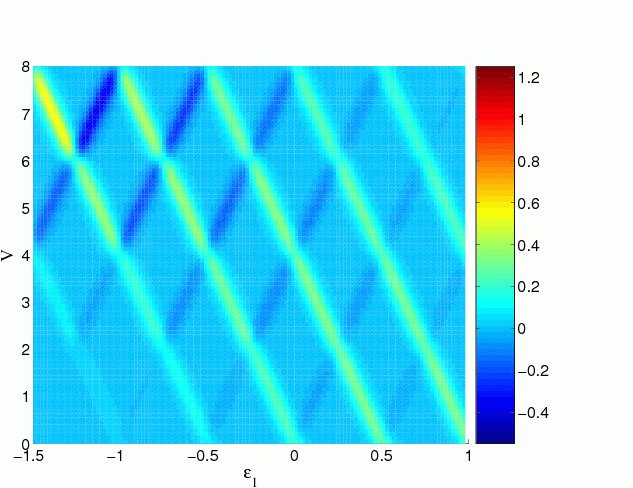}
\caption{The differential conductance, $dI/dV(\widetilde{\varepsilon}_1,V)$ for $g_1=2.5$, 
$g_2=0.1$, and $\Delta=\omega_0/2$ at $T=0.05\omega_0$.}\label{FIG.3}
\end{figure}

This peculiar feature can be understood from the bias-voltage-dependent electron occupation 
probabilities. For a system with $\widetilde{\varepsilon}_1=-0.25\omega_0$ [Fig.~4(a)], 
considering increasing bias voltage up to $0.5\omega_0$, the excited orbital (LUMO) starts to be 
occupied and then contributes to current. When the bias voltage arrives at $1.0\omega_0$, only 
two states, $1_0$ and $2_0$ ($i_n$ denotes the $n$-phonon-assisted tunneling channel of orbital 
$i$), are involved at the bias voltage window with respective symmetric tunneling rates, 
$x=\Gamma_{L1}^+=\Gamma_{R1}^-=\Gamma_{L1}{\cal L}_{0}^1\ll \Gamma_{L1}$ and 
$y=\Gamma_{L2}^+=\Gamma_{R2}^-=\Gamma_{L1}{\cal L}_{0}^2\approx \Gamma_{L1}$ (due to $g_2^2\ll 1 
\ll g_1^2$). This situation is equivalent to a two-level QD with symmetric tunnel-couplings under 
a large bias voltage at zero temperature, as indicated in Fig.~5(a). Thus, we have attained the 
maximum value of $\rho_{11}$ ($\rho_{11}\simeq\rho_{00}\simeq\rho_{22}\simeq 1/3$) and the 
current and Fano factor are given by 
\bn
I_a&=& \frac{x+y}{3}\Gamma_{L1}, \\
F_a&=& \frac{2(x^2+y^2)+xy}{9xy}, \label{Fa}
\en
respectively. From Eq.~(\ref{Fa}), we obtain the huge Fano factor (due to $x\ll y$), $F_a\simeq 
2y/9x\simeq 114$.  
If the bias voltage increases further, for example to $1.5\omega_0$, the tunneling channel $1_1$ 
is in alignment with the Fermi energy of the left lead. As a result, an electron in the QD 
prefers to re-occupy the ground orbital, which causes a suppression of current due to its 
stronger EPC constant than that of the excited orbital, $g_1\gg g_2$. In this case, the 
zero-temperature equivalent model is depicted in Fig.~5(b), where tunneling through the ground 
orbital becomes asymmetric, $x'=\Gamma_{L1}^+=\Gamma_{L1}({\cal L}_{0}^1+ {\cal L}_{1}^1) > 
x=\Gamma_{R1}^-$. Solving the simplified rate equations, one obtains
\bn
\rho_{11}&=& \frac{x'}{x'+2x}>\rho_{22}=\rho_{00}=\frac{x}{x'+2x}, \\
I_b &=& \frac{x(x'+y)}{x'+2x} \Gamma_{L1}< I_a, \\
F_b &=& \frac{2x'(y^2+x^2 -xy) + y(x'^{2}+2x^2)}{y(x'+2x)^2}.
\en        
It should be noted that if the bias voltage increases further to $2.5\omega_0<V<3.5\omega_0$, the 
system changes back to the symmetric model (a) with a stronger tunneling rate, 
$x=\Gamma_{L1}({\cal L}_{0}^1 + {\cal L}_1^1)$. 

\begin{figure}[htb]
\centering
\includegraphics[height=8cm,width=8.5cm]{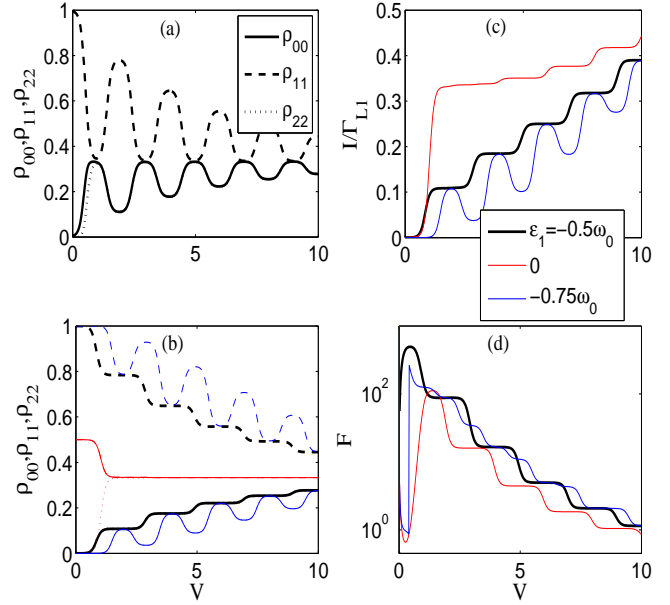}
\caption{(a,b) Electron occupation probabilities, (c) current $I$, and (d) Fano factor $F$ vs. 
the bias voltage for the two-level system with $\Delta/\omega_0=0.5$ and $g_1=2.5$. (a) is 
plotted for  $\widetilde\varepsilon_{1}/\omega_0=-0.25$; (b-c) are for 
$\widetilde\varepsilon_{1}/\omega_0=-0.5$ (thick lines), $0$ (thin red lines), and $-0.75$ (thin 
blue lines).}\label{FIG.4}
\end{figure}

\begin{figure}[htb]
\centering
\includegraphics[width=8.5cm]{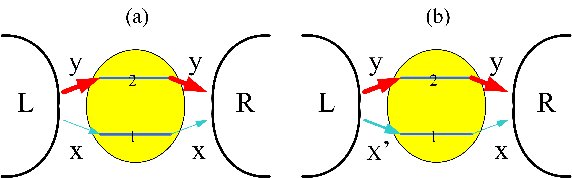}
\caption{Schematic diagram of two equivalent models for a two-orbital system. The arrows denote 
the directions of electron tunneling.}\label{FIG.5}
\end{figure}

Similar to the single-orbital QD, if the energy of the ground orbital is zero, 
$\widetilde{\varepsilon}_1=0$, the system always corresponds to the equivalent model (a). As a 
result, it exhibits step structures in the electron occupation probabilities [Fig.~4(b)] and 
current [Fig.~4(c)]. However, a system with $|\widetilde{\varepsilon}_1|\geq \omega_0/2$ is 
always equivalent to model (b): (1) if $|\widetilde{\varepsilon}_1|= n\omega_0/2$, electron 
occupation probabilities also exhibit step structures because of simultaneous increases of $x'$ 
and $x$ (Table I) with increasing bias voltage. Correspondingly, no NDC occurs in its current; 
(2) On the contrary, if $|\widetilde{\varepsilon}_1|\neq n\omega_0/2$, the tunneling-in rate $x'$ 
increases first with increasing bias voltage, which means that an electron has a greater 
opportunity to enter the ground orbital, i.e., $\rho_{11}$ increases. With further bias voltage 
increase, the tunneling-out rate $x$ also increases, leading to a decrease of $\rho_{11}$. 
Therefore, this system also exhibits oscillatory behavior in its electron occupation 
probabilities, and thus exhibits an NDC feature in its current.
The noise in all these cases is still super-Poissonian due to $x',x\ll y$ [Fig.~4(d)].

\begin{figure}[htb]
\centering
\includegraphics[height=7.5cm,width=8.5cm]{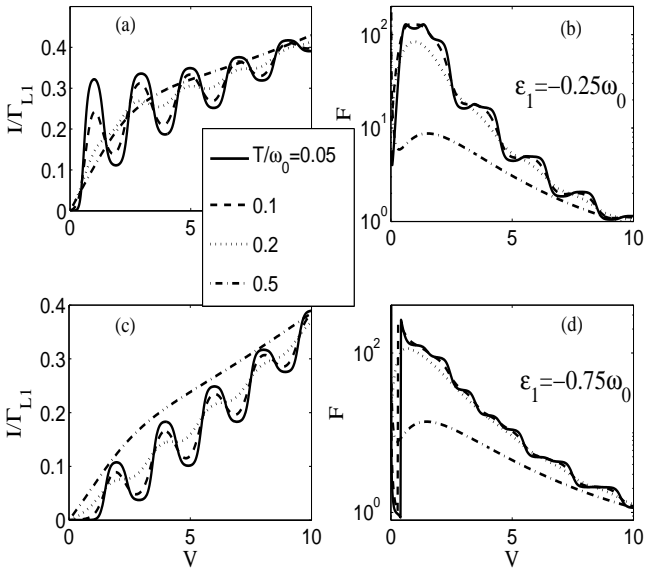}
\caption{Current $I$ (a,c) and Fano factor $F$ (b,d) vs. bias voltage for the two-level system 
with $\widetilde\varepsilon_{1}/\omega_0=-0.25$ (a,b) and 
$\widetilde\varepsilon_{1}/\omega_0=-0.75$ (c,d) for various temperatures $T$. Other parameters: 
$g_1=2.5$, $g_2=0.1$, and $\Delta/\omega_0=0.5$.}\label{FIG.4}
\end{figure}

Finally, we exhibit the temperature dependence of the NDC and super-Poissonian current noise for 
the two-level systems in Fig.~6. It is found that the NDC behavior of the current gradually 
becomes weaker with increasing temperature and nearly disappears at a relatively high 
temperature, $T=0.5\omega_0$, but the noise still remains strongly enhanced, $F\gg 1$, in the low 
bias-voltage region even though the Fano factor $F$ decreases with increasing temperature.     

\section{Conclusions}

In summary, we have analyzed equilibrated phonon-assisted tunneling and its zero-frequency 
current noise through a single molecular QD having two electronic orbitals with asymmetric 
couplings to the IVM using generic rate equations. Our results show that due to asymmetric EPC, 
the two-orbital molecular transistors exhibit strong and robust NDC and enhanced shot noise with 
a large Fano factor. We have also fully analyzed the reasons and conditions for the appearance of 
NDC in these systems.

\section*{Acknowledgment}

This work was supported by Projects of the National Science Foundation of China, the Shanghai 
Municipal Commission of Science and Technology, the Shanghai Pujiang Program, and Program for New 
Century Excellent Talents in University (NCET).

\end{document}